%% file: PIMA_prd.tex
\documentclass[%
	twocolumn,
	showpacs,
	amsmath,
	amssymb,
	aps,
	prd,
	nofootinbib
	]{revtex4}

\usepackage{xspace}
\usepackage{graphicx}
\usepackage{dcolumn}
\usepackage{MRFM_RevTex_macros}

\newcommand{\myecheck}{
\overset{\hspace*{0.4ex}%
\raisebox{-0.25ex}[0ex][0ex]{%
\includegraphics[width = 1.1ex]{PIMA_checkCharacter.epsf}}}%
{=}}

\newcommand{\myedef}{
\overset{\scriptscriptstyle%
\text{def}}{=}}%

\newlength{\myboxwidth}
\newlength{\mycaptionwidth}
\newcommand{\myStrut}{\rule[-1ex]{0pt}{4ex}}

\newenvironment{PIMAitemize}%
{%
\begin{list}{$\bullet$}{%
	\setlength{\topsep}{1ex}%
 	\setlength{\partopsep}{1ex}%
 	\setlength{\itemsep}{0.5ex}%
	\setlength{\leftmargin}{2\parindent}%
	\setlength{\rightmargin}{\parindent}%
	\setlength{\listparindent}{1.5\parindent}%
	\setlength{\parsep}{0.0 in}}}%
{\end{list}%
}

\newtheorem{definition}{Definition}
\newtheorem{conjecture}{Conjecture}
\newtheorem{lemma}{Lemma}

\begin{document}

\title{Path Integrals over Measurement Amplitudes: \protect\\ 
Practical Quantum Foundations for Signal Processing and Control}

\author{John A.~Sidles}
\altaffiliation[Address:\ ]{Department of Orthop{\ae}dics and Sports Medicine, Box 356500,
University of Washington, Seattle, Washington 98195}
\email[email:\ ]{sidles@u.washington.edu}
\homepage[home page:\ ]{http://faculty.washington.edu/sidles}
\affiliation{University of Washington}

\date{November 19, 2002}

\begin{abstract}
	It is shown that classical control diagrams can be mapped
	one-to-one onto quantum path integrals over measurement
	amplitudes.  To show the practical utility of this method,
	exact closed-form expressions are derived for the control
	dynamics and quantum noise levels of a test mass observed by
	a Fabry-Perot interferometer.  This formalism provides an
	efficient yet rigorous method for analyzing complex systems
	such as interferometric gravity wave detectors and magnetic
	resonance force microscopy (MRFM) experiments.  Quantum
	limits are conjectured for the sensitivity of
	interferometric observation of test mass trajectories.
\end{abstract}

\pacs{07.60.Ly, 07.79.Pk, 04.80.Nn, 95.55.Ym, 
03.65.Ta, 03.65.Ud, 03.67.Dd, 03.65.Xp, 42.50.Lc}

\maketitle

\section{Introduction}
\noindent As reviewed by Mensky \cite{Mensky:93}, the
formalism of path integrals over measurement amplitudes was
first suggested by Feynman \cite{Feynman:48} and was
subsequently worked out in greater detail by Mensky
\cite{Mensky:79a,Mensky:79b} and by Caves
\cite{Caves:86,Caves:87a}.

We show in this article that the formalism of path integrals
over measurement amplitudes can provide practical quantum
foundations for control theory, and we illustrate these
foundations by the worked example of resonant interferometric
observation of a test mass.

From a physics point of view we will work everything
backwards.  We will start, rather than finish, with a block
diagram that describes the dynamics of a classical system that
is subject to closed-loop control.  We will show that such
diagrams can be mapped one-to-one onto path integrals over
measurement amplitudes.  Then we will illustrate the physical
and control-theoretic significance of each path integral term
by analyzing a test mass observed by resonant optical
interferometry.  Finally, we will suggest that the dynamical
behavior of such systems is connected to unsolved problems in
quantum signal processing and cryptography.

\input{PIMA_figure_1.txt}

\section{Control Theory Foundations}
\noindent Engineers commonly formulate control theory in terms
of block diagrams and signal-flow graphs \cite{Kuo:91}.  A
block diagram is conceptually similar to a Feynman diagram: it
is a graphical representation of a set of equations.

Figure~\ref{fig:block diagram} shows a block diagram for a
test mass whose position $q(t)$ is continuously measured and
controlled.  The control force $f_{\text{c}}(t)$ is determined
from the commanded position $y_{\text{c}}(t)$ and the
estimated position $y(t)$ via kernels $\Gamma$ and
$\sprime\Gamma$:
\begin{subequations}
\begin{align}
	\label{eq:control a}\notag
   f_{\text{c}}(t) = & \int_{-\infty}^{t}\!\!\! d\sprime{t}\,\big[
	 \Gamma(t-\sprime{t}) (y_{\text{c}}(\sprime{t})-y(\sprime{t}))\big]\\
	 &-\int_{t}^{\infty}\!\!\! d\sprime{t}\,\sprime\Gamma(t-\sprime{t}) y(\sprime{t})\,.
\end{align}
Here by convention the feedback has been separated into a
control kernel $\Gamma$ that is causal and a backaction kernel
$\sprime \Gamma$ that is anticausal.

Anticausal backaction kernels are a natural idiom in control
theory.  To see why, consider a present fluctuation in
$y(\sprime t)$ that represents a photon detected at time
$\sprime t$, having bounced off the test mass at past times
$t$.  It follows that the backaction kernel
$\sprime\Gamma(t-\sprime t)$ \emph{must} be anticausal, in
order to describe the past-time force exerted by those
bounces.  There is no implication that the backaction physics
is noncausal.

To anticipate, all the other kernels of Fig.~\ref{fig:block
diagram} are explicitly causal, and so is the path integral
that we will derive for the system dynamics.  The overall
formalism is therefore explicitly causal, as we will discuss
following~(\ref{eq:fundamental}).

The controller kernel $\Gamma$ is the main subject of control
theory; it typically serves some useful purpose like moving
the test mass to a commanded position or stabilizing the
system dynamics.  Such kernels can be designed for optimal
performance~\cite{Garbini:96}, and essentially any desired
causal kernel can be realized by digital technology
\cite{Bruland:96}.

The estimated position $y(t)$ is determined from the test mass
position $q(t)$ and the measurement noise $q_{\text{n}}(t)$
via the measurement kernel $\xi$:
\begin{equation}
	\label{eq:control b}
   y(t) = \int_{-\infty}^{t}\!\!\! d\sprime{t}\,\xi(t-\sprime{t}) 
	\big(q(\sprime{t})+
	 q_{\text{n}}(\sprime{t})\big).\\
\end{equation}
It follows that $q(t)$ cannot be observed directly, but rather
must be estimated from $y(t)$; such estimation plays a central
role in control theory.

Finally, the dynamical behavior of $q(t)$ is determined by the
external force $f_{\text{e}}(t)$, the process noise
$f_{\text{n}}(t)$, and the control force $f_{\text{c}}(t)$ via
the process kernel $G$:
\begin{equation}
	\label{eq:control c}
	q(t) = \int_{-\infty}^{t}\!\!\! d\sprime{t}\,G(t-\sprime{t}) 
	\big(f_{\text{e}}(\sprime{t})+
	 f_{\text{n}}(\sprime{t})+f_{\text{c}}(\sprime{t})\big).
\end{equation}
\end{subequations}

Our main statistical assumption is that $q_{\text{n}}(t)$ and
$f_{\text{n}}(t)$ are stationary zero-mean random processes;
we will show that path integrals naturally generate quantum
noise with this property.  Without loss of generality, we can
further specify that $q_{\text{n}}(t)$ and $f_{\text{n}}(t)$
are statistically independent; for any given block diagram
this can be arranged by ``pulling'' correlated noise through
$G$ and suitably redefining $q_{\text{n}}(t)$ and
$f_{\text{n}}(t)$.

Then the Fourier-domain solution to \mbox{(\ref{eq:control
a}--c)} is such that $\stilde{y}(\omega)$ has mean value
\begin{subequations}
\begin{equation}
	\label{path:a}
	\expect{\stilde{y}(\omega)} = \frac{\stilde{\xi}(\omega)
	\stilde{G}(\omega)\,(
	\stilde{f}_{\text{e}}(\omega) + 
	\stilde{\Gamma}(\omega)\stilde{y}_{\text{c}}(\omega))}
	{1+\stilde{\xi}(\omega)\stilde{G}(\omega)
	\big(\stilde{\Gamma}(\omega)+\sprime{\stilde{\Gamma}}(\omega)\big)}
\end{equation}
and spectral density
\begin{align}
	\label{path:b}
	S_{y}(\omega)& = \frac{|\stilde{\xi}(\omega)|^{2}\,\big(S_{q}(\omega) + 
	|\stilde{G}(\omega)|^{2}S_{f}(\omega)\big)}
	{\big|1+\stilde{\xi}(\omega)\stilde{G}(\omega)
	\big(\stilde{\Gamma}(\omega)+\sprime{\stilde{\Gamma}}(\omega)\big)\big|^{2}}.
\end{align}
\end{subequations}
The form of this result reflects our convention that the
measurement noise $q_{\text{n}}(t)$ and the process noise
$f_{\text{n}}(t)$ are uncorrelated.  Here our notation and
normalization convention for Fourier transforms and spectral
densities is
\begin{subequations}
\begin{align}
	\label{eq:spectraldensity-a}
S_{f}(\omega) & \myedef
\int_{-\infty}^{\infty}d\tau\,\expect{f(0)f(\tau)}e^{-i\omega\tau},\\
	\label{eq:spectraldensity-b}
\stilde{f}(\omega) & \myedef
\int_{-\infty}^{\infty}d\tau\,f(\tau)e^{-i\omega\tau},\\
	\label{eq:spectraldensity-c}
\stilde{\psi}(\omega,\sprime{\omega}) & \myedef
\int_{-\infty}^{\infty}d\tau
d\sprime{\tau}\,\psi(\tau,\tau')e^{-i(\omega\tau+\sprime\omega\sprime\tau)},
\end{align}
\end{subequations}
with $\expect{\ldots}$ designating an ensemble average.  Thus
our spectral densities are ``two-sided.''  We regard the
kernels of \mbox{(\ref{eq:control a}--c)} as defined for all
times $\tau$ with $\Gamma(\tau)=\xi(\tau)=G(\tau)=0$ for
$\tau<0$ and $\sprime\Gamma(\tau)=0$ for $\tau>0$.

As a mathematical point, the functions $f_{\text{n}}(t)$ and
$q_{\text{n}}(t)$ in the block diagram of Fig.~\ref{fig:block
diagram} do \emph{not} appear as independent functions in
(\ref{path:a}--b).  Neither will these functions appear as
independent variables in our path integrals---not even as
dummy variables of integration---nor will they appear in any
subsequent part of our formalism.  Their sole role is as
mnemonic aids: they remind us to include $S_{f}$ and $S_{q}$
in~(\ref{path:b}).

Since $f_{\text{n}}(t)$ and $q_{\text{n}}(t)$ do not appear in
our formalism as indendent functions, we have no mathematical
or physical basis for assigning independent meanings to them. 
We can only speak of them in terms of a unitary equivalent
noise having, \emph{e.g.}, spectral density $S_{f}(\omega) +
|\stilde{G}(\omega)|^{-2}S_{q}(\omega)$ at the $G$ block
input, per Fig.~\ref{fig:block diagram}.  Again anticipating
future results, our sole motivation for maintaining
$S_{f}(\omega)$ and $S_{q}(\omega)$ as separate densities is
to express the noise reciprocity relation
(\ref{eq:reciprocity}) in a device-independent form.

This unitary point of view is consonant with information
theory, since in light of the above discussion it is not
possible---even in principle---to infer independent values for
$f_{\text{n}}(t)$ and $q_{\text{n}}(t)$ from the measured
quantity $y(t)$.  Furthermore, maintaining a unitary point of
view will forestall conceptual difficulties in
Section~\ref{sec: literature}, where we compare path integral
results with analyses of shot noise and radiation-pressure
noise in the literature.

\section{Path Integral Representations \protect\\ of Control Theory}

\noindent Now we seek a path integral that reproduces
((\ref{path:a}--b)).  We approach this as a purely mathematical
exercise whose sole requirements are tractability and
generality.  Adopting the path integral notation of Brown
\cite{Brown:92}, we consider functionals of the form
\begin{subequations}
\begin{align}
	\label{eq: path integral}
	P(y(t)&|f_{\text{e}}(t), y_{\text{c}}(t)) = \notag\\
	&\left|\int [dq]
	\exp\big[\lcal{A}(q(t),y(t),f_{\text{e}}(t),y_{\text{c}}(t))\big]\right|^{2}.
\end{align}
Here $P(y(t)|f_{\text{e}}(t), y_{\text{c}}(t))$ is a Gaussian
probability functional whose mean and variance must reproduce
(\ref{path:a}--b).

\input{PIMA_table_1.txt}

For $P$ to be of the required Gaussian form, the action
functional $\lcal{A}$ must be biquadratic in $q(t)$ and $y(t)$
and bilinear in $f_{\text{e}}(t)$ and $y_{\text{c}}(t)$.  Note
that $q(t)$ appears only as a dummy variable of integration
that is not yet identified as the test mass trajectory of
Fig.~\ref{fig:block diagram}.
Adopting a conventional form that facilitates subsequent connection to
control theory, the most general functional $\lcal{A}$ that reproduces
(\ref{path:a}--b) can be written as
\begin{align}
	\lcal{A} & =  \int_{-\infty}^{\infty}\frac{dt}{i\hbar}\,
	\big[\,-\lcal{L} & \text{Lagrangian action}\notag\\
	& \quad +\lcal{H}_{\text{f}} + \lcal{H}_{\Gamma}  & \text{external force and control}\notag\\
	& \quad+ \lcal{M}_{\sprime\Gamma} + \lcal{M}_{\theta} +  \lcal{M}_{\psi} \hspace*{-12em}
	& \text{backaction effects}\notag\\
	& \quad+ i \lcal{M}_{\xi}\,\big]  & \text{measurement amplitude}
	\label{eq: path integral terms}
\end{align}
\end{subequations}
where $\{\lcal{L}, \lcal{H}_{\text{f}}, \lcal{H}_{\Gamma},
\lcal{M}_{\sprime\Gamma}, \lcal{M}_{\theta},
\lcal{M}_{\psi},\lcal{M}_{\xi}\}$ are real-valued functionals
that are given explicitly in Table~\ref{tab:functionals} in
terms of measurement kernels $\{\sprime\Gamma, \Gamma, \theta,
\psi, \xi\}$.  Carrying through the path integral by methods
that are essentially algebraic \cite{Brown:92}, we connect the
kernels of (\ref{eq: path integral terms}) to the control
dynamics of (\ref{path:a}--b), as summarized in
Table~\ref{tab:connections}.

This completes our goal of establishing a path integral
representation of the control diagram of Fig.~\ref{fig:block
diagram}. 

\input{PIMA_table_2.txt}

\section{The Measurement Amplitude \mbox{For Optical Interferometry}}
\noindent To show what is gained by attacking the problem in
this systematic way, we now calculate the measurement
amplitude kernels $\{\sprime\Gamma, \Gamma, \theta, \psi,
\xi\}$ for a resonant optical interferometer.  Then we 
systematically read off the system dynamics and quantum noise
from Tables~\ref{tab:functionals} and~\ref{tab:connections}
and Eqs.~(\ref{path:a}--b).

To carry through this calculation---and indeed to carry
through \emph{any} path integral/measurement amplitude
calculation---it suffices to specify the classical optical
scattering amplitude and the photon detection statistics. 
From classical physics we know that for a general single-port
optical device the outgoing amplitude $a_{\text{out}}$ at time
$t$ is causally conditioned upon the past history of the
internal coordinate $q(t)$ as follows:
	\begin{align}
		\label{eq:classical scattering amplitude}
	&  a_{\text{out}}(t|q(t)) = a_{\text{in}} e^{i\zeta}\,\bigg(1 + \int_{-\infty}^{t}\!\!d\sprime{t}\,
   \alpha(t-\sprime{t}) q(\sprime{t}) \notag \\
	&  \qquad +  \int_{-\infty}^{t}\!\!d\sprime{t}\int_{-\infty}^{t}\!\!d\spprime{t}\,
   \beta(t-\sprime{t},t-\spprime{t}) q(\sprime{t}) q(\spprime{t}) \bigg).
	\end{align}
Here $a_{\text{in}}$ is the input light amplitude (assumed
constant), $\alpha$ and $\beta$ are scattering kernels, and
$e^{i\zeta}$ is an overall phase.  Perturbations of order
$q^{3}$ and higher are neglected, and by convention we
normalize $a_{\text{in}}$ such that the photon input rate is
$r_{\text{in}} = |a_{\text{in}}|^{2}$.  Causal boundary
conditions are imposed.  Then the mean rate $\expect{r(t)}$ at
which photons are detected at time $t$ is a functional of the
past trajectory $q(t)$: $\expect{r(t)} =
|a_{\text{out}}(t|q(t))|^2$.

Now we are ready for the key element of our
formalism.  We introduce as an \emph{ansatz} the following
fundamental relation between the quantum measurement action of
\mbox{(\ref{eq: path integral}--b)} and the classical
scattering amplitude (\ref{eq:classical scattering
amplitude}):
\begin{align}
	\label{eq:fundamental}
\exp\left[
\frac{\lcal{M}(r(t),q(t))}{i\hbar}\right]
& =  \left[\frac{a_{\text{out}}(t|q(t))}
{|a_{\text{out}}(t|q(t))|}\right]^{r(t)}\notag\\
& \hspace{-3em} \times
\exp\left[
\frac{-(r(t)-|a_{\text{out}}(t|q(t))|^2)^2}
{4 \alpha_{\text{s}}|a_{\text{out}}(t|q(t))|^2}\right].
\end{align}
Here $\lcal{M}(r(t),q(t)) \myedef \lcal{M}_{\sprime\Gamma} + 
\lcal{M}_{\theta} +  \lcal{M}_{\psi} + i \lcal{M}_{\xi}$
is the  measurement action that appears in (\ref{eq: path integral}--b).

We will present no field-theoretic justification for this
\emph{ansatz}, and in Section~\ref{sec:Discussion} we will
present reasons for thinking that a rigorous field-theoretic
justification would involve deep quantum-informatic issues. 
Instead, our limited goal in this article will be to show that
the \emph{ansatz} reproduces, within a path integral
formalism, known classical and quantum physics.

From the \emph{ansatz} we immediately obtain rules for
translating the optical kernels $\alpha$ and $\beta$ into the
measurement amplitude $\lcal{M}$ of (\ref{eq:fundamental}). 
No physical insight is employed; instead we impose the purely
algebraic requirement that the rules convert
(\ref{eq:fundamental}) into an identity (up to
$\lcal{O}(q^{2})$ in the path integral functional).  The rules
are summarized in Table~\ref{tab:sidebandrules}.
Because the optical kernels $\alpha$ and $\beta$ satisfy
causal boundary conditions, the measurement amplitude
(\ref{eq:fundamental}) and the resulting path integral
(\ref{eq: path integral}) are explicitly causal as promised
in the discussion following (\ref{eq:control a}).

\input{PIMA_table_3.txt}

Non-rigorously, the \emph{ansatz} can be derived by
constructing a numerical wave function simulation along the
lines given by Gardiner and Zoller \cite{Gardiner:00},
with each photon detected separately and accounted
numerically.  Such simulations become exponentially slower as
the number of photons and the multiple reflections of each
photon are increased; this illustrates the well-known
nonpolynomial difficulty of quantum simulation in general. 
Coding such simulations with a view to making them numerically
efficient leads naturally to a path integral formalism.  This
is the path the author followed to the results of this
article.

The \emph{ansatz} embodies two key physical principles, which
are both well satisfied in optical interferometry.  First,
individual photons are assumed to be detected discretely, such
that the decoherence associated with each detection event
procedes to completion within a very short time compared to
all other dynamical time scales of the system.  Under these
circumstances a well-defined phase can be associated with
photon detection.  The interference of these phases can be
readily observed, which is of course the reason such
measurements are called ``interferometric''.  The
\emph{ansatz} functional
$\left(a_{\text{out}}/|a_{\text{out}}|\right)^{r(t)}$
accumulates the test mass phase from repeated photon
detections; this phase creates the quantum backaction.

Second, it is assumed that large numbers of photons are
detected, in which case the detection statistics can
reasonably be described by a counting formula of the usual
Gaussian form (the lower right-hand term in
(\ref{eq:fundamental})), such that the photon flux spectral
density is \mbox{$S_r = \alpha_{\text{s}}r_{\text{in}}$}, with
due allowance for photon number squeezing as parametrized by
$\alpha_{\text{s}}$.  The physical role of the Gaussian term
is to restrict the domain of path integration, conditioned
upon the flux measurement $r(t)$, in precisely the manner
envisioned by Feynman, Mensky, and Caves.

%
%

\section{A Worked Example:\protect\\ Single-Port Fabry-Perot Interferomtry}
\label{sec:Example}
\noindent To make the path integral/measurement amplitude
formalism come alive we will apply it to an engineering
analysis of the Fabry-Perot cavity shown in Fig.~\ref{fig:PIMA
interferometer}.  This is a single-arm interferometer with
single-port detection; it is not intended to represent a
realistic gravity wave detector.  However, even this simple
design exhibits complex dynamical and noise phenomena; our
goal is to show how to use path integral methods in analyzing
this behavior.

Many of the results that we will obtain by path integration
have also recently been obtained by operator methods in the
literature on gravity wave detection; this literature is
reviewed in Section~\ref{sec: literature}.  We will find no
serious conflict between operator methods and the path
integral/measurement amplitude formalism.

In applying the path integral formalism, our sole
computational job is to calculate the optical kernels
$\stilde{\alpha}(\omega)$ and $\stilde{\beta}(\omega,-\omega)$
of (\ref{eq:classical scattering amplitude}) for the
Fabry-Perot cavity of Fig.~\ref{fig:PIMA interferometer}; the
rest is substitution into the rules of
Tables~\ref{tab:connections} and~\ref{tab:sidebandrules}.

\input{PIMA_figure_2.txt}

The input light is right-going, as shown in Fig.~\ref{fig:PIMA
interferometer}, and our phase convention is that it has
space-time dependence $a(x,t) = a_{\text{in}} e^{i k x -
\omega_{0}t}$, where the wave number $k$ and the optical
carrier freqency $\omega_{0}$ are positive quantities.  This
accords with the quantum mechanics phase convention that a
right-going quantum carrying momentum $\hbar k$ has a spatial
wave function $\propto e^{i k x}$.  With this convention,
positive mirror displacements $q(t)$ correspond to longer
cavity lengths, as shown in Fig.~\ref{fig:PIMA
interferometer}, and positive forces act to push the mirrors
apart.

This same phase convention, when conjoined with the Fourier
convention (\ref{eq:spectraldensity-b}), prescribes that
optical sidebands at a frequency $\omega_{\text{m}}$ have a
Doppler-shifted time-dependence \mbox{$e^{-i \omega_{0}t +
i\omega_{\text{m}} t}$}.  Positive-frequency sidebands
($\omega_{\text{m}} >0$) are therefore associated with
\emph{redshifted} optical quanta.  This unintuitive convention
will be important later on when we check energy conservation.

Inspection of Table~\ref{tab:sidebandrules} shows that the
dynamical and noise behavior of the system is completely
determined by the Fabry-Perot sideband amplitude
$\stilde\alpha(\omega)$ and carrier amplitude
$\stilde\beta(\omega,-\omega)$.  A straightforward 
perturbative calculation yields for the sideband
\begin{subequations}
\begin{align}
	\label{eq:FPa}
	 & \stilde\alpha(\omega)  =\frac{
	 2 i k e^{i(3 \omega\tau-2\phi)}\sin^{2}\rho}
	 {(e^{2i(\omega\tau-\phi)}-\cos\rho)(1-e^{-2i\phi}\cos\rho)},\\
\intertext{and for the carrier}
	\label{eq:FPb}
	 & \frac{\stilde\beta(\omega,-\omega)}{
	 \stilde\alpha(\omega)\stilde\alpha(-\omega)} = \frac
	 {\sin^{2}\rho+2 i \sin(2\phi)\cos\rho} {2\sin^{2}\rho}.
\end{align}
\end{subequations}
Here the power reflectivity of the input mirror is by definition
$\cos^{2}\rho$, and the one-way optical phase length $\phi$ of
the cavity is $\phi = k L + \pi$, with $L$ the cavity length. 
The extra $\pi$ in the definition of $\phi$ is conventional;
it ensures that tuning to $\phi = 0 \pmod{2\pi}$ yields
maximal intracavity optical power.  For fixed $\phi$, and
therefore fixed optical power, we can maximize the sideband
amplitude by tuning the mirror modulation frequency $\omega$
to $\omega \tau = \phi \pmod{2\pi}$; this is
physically equivalent to tuning the sideband on-resonance.

The same perturbative calculation yields for the phase
$e^{i\zeta}$ of the output light
\begin{equation}
	e^{i\zeta} = \frac{e^{-2i\phi}\cos\rho-1}{\cos\rho-e^{- 2 i \phi}},
\end{equation}
and it is easy to check that $|e^{i\zeta}|=1$.
It is also useful to know the cavity power gain
$\lcal{G}(\phi)$:
\begin{equation}
	\label{eq:cavityGain}
	 \lcal{G}(\phi) \myedef \frac{\text{cavity power}}{\text{input power}} 
	 = \left|\frac{\sin\rho}{\cos\rho-e^{- 2 i \phi}}\right|^{2}.
\end{equation}
$\lcal{G}(\phi)$ is a period-$\pi$ function of
$\phi$, peaked about $\phi\sim 0$, and having
half-width-half-maximum $\phi_{\lcal{F}}$
\begin{equation}
	\label{eq:finesse phase}
	\phi_{\lcal{F}} = \frac{1}{2}\,\sqrt{\cos\rho + \cos^{-1}\rho -2}.
\end{equation}
By construction, $\phi_{\lcal{F}}$ is the extra phase length
required to reduce the on-resonance cavity power by 1/2. 
Conventionally $\lcal{F}\equiv \pi/(2\phi_{\lcal{F}})$
is called the \emph{finesse} of the cavity, and for
high-finesse cavities the on-resonance intracavity power gain
is $\lcal{G}(0) \simeq 2\lcal{F}/\pi=\phi_{\lcal{F}}^{-1}$. 
We will always specify changes in cavity length as multiples
of $\phi_{\lcal{F}}$.

Now we are ready to specify an interferometer design. 
We choose length scales and power levels that are
characteristic of recent proposals for advanced gravity-wave
detectors \cite{LIGO:02}:
\begin{center}
\begin{tabular}{r@{:\ }rl}
cavity length& $L$&$=4\ \text{km}$\\
detected power& $P_{\text{out}}$&$=180\ \text{W}$\\
on-resonance power& $P_{\text{max}}$&$=830\ \text{kW}$\\
light wavelength& $\lambda$&$=1064\ \text{nm}$\\
test mass& $m$&$=40\ \text{kg}$
\end{tabular}
\end{center}
This is a high-finesse cavity, with $\lcal{F} \simeq 7240$,
corresponding to a mirror power reflectivity $\cos^2\rho \simeq
0.99913$.

\input{PIMA_figure_3.txt}

We begin by considering the static behavior of the system. 
The path integral prediction for the light force
${f}_{\text{e}}(t)$ on the mirror can be read off from
Tables~\ref{tab:connections} and ~\ref{tab:sidebandrules}:
\begin{align}
	{f}_{\text{e}}(t) & \myedef {\textstyle
	\int_{-\infty}^{\infty}\!\!\frac{d\omega}{2\pi}}\,e^{i\omega t}\stilde{f}_{\text{e}}(\omega)
	&\text{\hspace*{-3em} by definition}\notag\\
		& = {\textstyle\int_{-\infty}^{\infty}\!\!\frac{d\omega}{2\pi}}\,
		\stilde{b}(\omega) \stilde{\theta}(-\omega) 
		& \text{by Table~\ref{tab:connections}}\notag\\
		& = {\textstyle\int_{-\infty}^{\infty}\!\!\frac{d\omega}{2\pi}}\,
		(-2 \pi r_{\text{in}}) \delta(\omega) \sprime{\stilde{\Gamma}}(\omega)
		& \text{by Table~\ref{tab:sidebandrules}}\notag\\
		& = -r_{\text{in}} \sprime{\stilde{\Gamma}}(0)
		& \text{by evaluation}	\notag\\
		& =   r_{\text{in}}\,\frac{\hbar}{2 i}
	 \big[\stilde{\alpha}(0)-\stilde{\alpha}^{\star}(0)\big]
	 &  \text{by Table~\ref{tab:sidebandrules}}\notag
\end{align}
By explicit calculation we check that this accords with
the standard expression for light pressure: 
\begin{equation}
{f}_{\text{e}}(t)
\myecheck 2 \hbar k r_{\text{in}}\lcal{G}(\phi).
\end{equation}
Here $2 \hbar k$ is the momentum transferred to the mirror by
the reflection of a single photon and
$r_{\text{in}}\lcal{G}(\phi)$ is the flux of photons incident
on the mirror.

Similarly, from Table~\ref{tab:connections} we see that the
zero-frequency spring constant exerted by the light is
$k_{\text{spring}}(\phi) \myedef \lim_{\omega\to
0}\stilde\psi(\omega,-\omega)$, and by explicit calculation we
check that this accords with the spring constant predicted
from the cavity gain $\lcal{G}(\phi)$:
\begin{equation}
k_{\text{spring}}(\phi) 
\myecheck -2 \hbar k^{2} r_{\text{in}}\frac{
\partial\lcal{G}(\phi)}{\partial\phi}.
\end{equation}
As a final check on the static physics, it is straightforward
to show that the large $\lcal{F}$ limit of
$k_{\text{spring}}(\phi)$ precisely accords with the Fabry-Perot
spring constant calculated by Braginsky, Khalili, and Volikov
\cite{Braginsky:01c} from classical physics.  We thus confirm
that the path integral/measurement amplitude method accurately
reproduces known results relating to static optical forces and
springs in resonant cavities.

In subsequent calculations we will not show all the steps, but
our results are always obtained by a similarly direct
application of the measurement amplitude rules given in
Tables~\ref{tab:connections}--\ref{tab:sidebandrules}.  These
rules are readily processed by symbolic programs; this reduces
the incidence of algebraic error.  Analytic continuation is
straightforward because the kernels are given in closed form;
we will see that this simplifies stability analysis.

Now we turn our attention to the practical challenge of tuning
the interferometer for dynamical stability and good noise
performance.  The static optical force and spring constant are
shown in Fig.~\ref{fig:statics}.  The optical forces are
weak---a few millinewtons at most---but the spring constant
can approach 100~N/$\mu$m, which is extraordinarily stiff. 
For an optical beam of nominal diameter 20~cm and length 4~km
the stiffness is equivalent to a modulus \mbox{$\sim
12.3$~TPa}, which is twelve times stiffer than an equivalent
bar of diamond.

This illustrates that light itself can serve as a structural
material, as was first recognized and explored for design
purposes by Braginsky, Gorodetsky and Khalili
\cite{Braginsky:97, Braginsky:99, Khalili:01} and subsequently
by Buonanno and Chen \cite{Buonanno:01a,Buonanno:02b,
Buonanno:02a}.

By inspection of Fig.~\ref{fig:statics} we see that
static stability is possible if and only if the interferometer
is tuned ``long'' (\emph{i.e.}, cavity phase length $\phi>0$),
and we henceforth confine our attention to this range.  In
practice, tuning is achieved by applying a few mN of force to
the mirror to press it against the optical spring; the
magnitude of the static force determines the equilibrium
cavity length and therefore the optical tuning.

Static stability does not guarantee dynamic stability, and on
physical grounds we expect Fabry-Perot cavities to be
dynamically unstable.  We reason as follows: if we push the
mirror against the optical spring, the spring will push back,
but only after a time lag while the intracavity intensity
builds up.  By oscillating the mirror, we can continuously
extract energy from the system.  

Physically, the possibility of energy extraction indicates the
presence of dynamical instability.  We will now prove that
such instabilities exist---for all cavity tunings---by
calculating the transfer function of the system in closed
analytic form.

We begin by noting that the time-averaged flux of output
photons from any linear lossless optical device must equal the
flux of input photons.  It is easy to check that the
Fabry-Perot amplitudes \mbox{(\ref{eq:FPa}--b)} satisfy this
constraint, which requires that $\{\alpha,\beta\}$ satisfy
\begin{align}
	\label{eq:flux conservation}
	|\stilde{\alpha}(\omega)|^{2} +  
	|\stilde{\alpha}(-\omega)|^{2} &	
	\myecheck  -2 * \text{Re}\big[\stilde{\beta}(\omega,-\omega)\big] ,
	\\[1ex]
	\text{[sideband photon flux]} & =  
	\text{[decreased carrier flux]}. \notag
\end{align}
Physically, photons that disappear from the carrier must
reappear in the sidebands.

This does not guarantee energy conservation, since the
outgoing sideband photons are Doppler shifted per the
discussion preceding (\ref{eq:FPa}--b).  To check energy
conservation we apply an external force $f_{\text{e}}(t)$ to
the mirror, such that the mirror is driven at an amplitude
$q(t) = q_{0}\cos(\omega_{\text{m}} t)$.  Then mechanical
energy is supplied to the mirror at a rate given by
(\ref{path:a}) as
\begin{align}
 \expect{f_{\text{e}}(t)\sdot{q}(t)}_{t}
 =i\omega_{\text{m}} \frac{q_{0}^{2}}{4}\, &
 \left[
 \sprime{\stilde{\Gamma}}(-\omega_{\text{m}})\stilde\xi(-\omega_{\text{m}})
 \right.
 \notag\\
& \left.\qquad -\,\sprime{\stilde{\Gamma}}(\omega_{\text{m}})\stilde\xi(\omega_{\text{m}})
 \right]
\end{align}
with
$\expect{\ldots}_{t}$ denoting a time average.
After substitution from Tables~\ref{tab:connections} and~\ref{tab:sidebandrules} this is
\begin{equation}
	\label{eq:energy conservation}
 \expect{f_{\text{e}}(t)\sdot{q}(t)}_{t} \myecheck \hbar \omega_{\text{m}} r_{\text{in}} \frac{q_{0}^{2}}{4}
 \left[|\stilde{\alpha}(-\omega_{\text{m}})|^{2} -  
	|\stilde{\alpha}(\omega_{\text{m}})|^{2}\right].
\end{equation}
Taking into account the Doppler sign convention discussed at
the start of this section, we recognize this as precisely the
excess optical power emitted in the sidebands.  Thus the
optomechanical instability is energetically driven by Doppler
shifts in the sidebands, such that energy is explicitly conserved
overall.

Now we analyze the instability in detail.  From
(\ref{path:a}), the transfer function $\stilde{T}(\omega)$ for
the cavity is
\begin{equation}
	\stilde{T}(\omega) \myedef
\hbar k c\,\frac{\stilde{y}(\omega)}{\stilde{f}_{\text{e}}(\omega)}
= 	\hbar k c\,\frac{\stilde{\xi}(\omega)
	\stilde{G}(\omega)}
	{1+\stilde{\xi}(\omega)\stilde{G}(\omega)\sprime{\stilde{\Gamma}}(\omega)}
\end{equation}
where the factor $\hbar k c$ normalizes $\stilde{T}(\omega)$
to units of watts of detected optical power per newton of
applied force.  

\input{PIMA_figure_4.txt}

In control theory---where Laplace transforms are more common
than Fourier transforms---it is standard practice to plot the
dominant poles and zeros of the transfer function in the
complex $s$-plane, with $s\myedef i \omega$. 

The path integral/measurement amplitude formalism gives
$\stilde{T}(\omega)$ in analytic form, and when such forms are
available, a standard technique in control theory is to
calculate a Pad\'{e} approximant to $\stilde{T}(\omega)$ using
a symbol manipulation program.  Carrying through this
calculation, we find that a~$\{1,4\}$ approximant yields a
good fit, and is comprised by a fixed zero at $s=0$ and four
dominant poles whose trajectories are shown in
Fig.~\ref{fig:PIMA_bode}.

The fixed zero at $s=0$ means that the cavity has zero static
sensitivity at all tunings.  Physically, this means that at
zero frequency the measured output photon flux equals the
input flux, no matter what the cavity length, as enforced by
photon conservation (\ref{eq:flux conservation}).

Dynamically, the cavity is optomechanically unstable at all
tunings, with the strongest instability at
\mbox{$\phi/\phi_{\lcal{F}} \sim 1$}.  To stabilize the
cavity, an additional control kernel $\stilde\Gamma(\omega)$
must be added.  In principle, a perfectly linear and noiseless
control kernel will not alter the signal-to-noise ratio
\cite{Garbini:96,Bruland:96}---because control kernels affect
signal and noise equally---but nonetheless it is good
engineering practice to choose a cavity tuning such that the
control challenges are not too great.

We choose a far-off-resonance cavity tuning
$\phi_{\text{tune}}=10\,\phi_{\lcal{F}}$.  Per
(\ref{eq:cavityGain}), this tuning reduces the cavity optical
power from the peak power of 830~kW to
8.2~kW---a~99\%~reduction.  In the ordinary course of events,
we might expect such low power to greatly diminish the
sensitivity of the interferometer.

However, the design compensates for the low cavity power by
exploiting two mitigating factors.  First, on physical grounds
we expect that the signal sideband will be resonant with the
cavity, and therefore passively amplified, at a frequency $
\omega_{\text{optical}}$ determined by (\ref{eq:FPa}) to be
$\omega_{\text{optical}}/(2\pi) \sim
\phi_{\text{tune}}/(2\pi\tau) \simeq 25.9~\text{Hz}$. 
Second, we expect the system to exhibit a mechanical resonance
at a frequency $\omega_{\text{mech}}$ such that
$\stilde\psi(\omega_{\text{mech}},-\omega_{\text{mech}})
\simeq m\omega_{\text{mech}}^{2}$, \emph{i.e.}, at a frequency
determined by the strength of the optical spring, and we
further expect this mechanical resonance to be unstable.  For
$\phi_{\text{tune}}=10\,\phi_{\lcal{F}}$ a numerical analysis
predicts this resonance at $\omega_{\text{mech}}/(2\pi)\simeq
23.7~\text{Hz}$.

Thus, on physical grounds we expect the transfer function to
exhibit one stable optical pole and one unstable mechanical
pole, both at about 20~Hz.

These expectations are in excellent accord with the Pad\'{e}
analysis shown in Fig.~\ref{fig:PIMA_bode}, which at
$\phi=10\,\phi_{\lcal{F}}$ exhibits a stable pole at
21.2~Hz with a quality $Q = 1.90$ and an unstable pole at
15.9~Hz with $Q = -2.59$.  These resonant frequencies are
slightly reduced relative to the above rule-of-thumb
expectations; this presumably reflects damping effects (which
characteristically lower resonant frequencies) combined with
the optomechanical coupling generated by the backaction kernel
$\sprime{\stilde\Gamma}$.

Bearing in mind the consensus view of control theorists that
``generally speaking, an unstable system is considered to be
useless'' \cite{Kuo:91}, these results provide a well-posed
starting point for addressing important practical questions
such as: is this Fabry-Perot system ``observable'' and
``controllable'' in the rigorous sense that these terms have
in control theory?  And if so, what would be a suitable design
for the control kernel $\stilde\Gamma(\omega)$?

We note that in control theory---and in any continuous
measurement theory---there is no sharp distinction between a
``position meter'' and a ``velocity meter.''  If position is
observable, then so is velocity, via a differentiating
filter.  Conversely, if velocity is observable, then so is
position, via an integrating filter.

We will not consider these control issues further because an
article-length exposition would be required, and because they
are a standard topic in control engineering textbooks
\cite{Kuo:91}.  Instead, we will simply assume that a
stabilizing controller is present, we will further assume that
it contributes negligible noise, and we will proceed to
analyze the sensitivity of the interferometer.

In keeping with accepted practice of the gravity-wave
community, we conflate all noise sources into single
equivalent force noise having spectral density
$S_{f}^{\text{tot}}(\omega)$, and we express this net force
noise as an equivalent strain noise
$S_{h}^{\text{tot}}(\omega)$ according to
\begin{equation}
	\label{eq: strain noise}
	S_{h}^{\text{tot}}(\omega) \equiv
	\frac{S_{f}^{\text{tot}}(\omega)}{(m\omega^{2} L)^{2}}
\end{equation}
where $L$ is the arm length and $m$ is the test mass. 
Physically, this convention acknowledges that audio-frequency
gravity waves, when observed over kilometer length scales, are
dynamically equivalent to tidal forces.

Combining this convention with (\ref{path:b}), we obtain for
the total equivalent strain noise
\begin{align}
	\label{eq:total strain noise}
S_{h}^{\text{tot}}(\omega) & = 
	\frac{S_{f}(\omega)}{(m\omega^{2} L)^{2}}
	+ \frac{S_{q}(\omega)}{(m\omega^{2} L)^{2}|\stilde G(\omega)|^{2}}\\
	& =\ \left[\begin{matrix}
		\text{process}\\
		\text{noise}
	\end{matrix}\right]
	+ \left[
	\begin{matrix}
		\text{measurement}\\
		\text{noise}
	\end{matrix}
		\right]\notag
\end{align}
where the functional forms of $S_{f}(\omega)$,
$S_{q}(\omega)$, and $\stilde G(\omega)$ are given in
Tables~\ref{tab:connections} and~\ref{tab:sidebandrules}.  The
feedback kernels $\{\sprime{\stilde\Gamma}, \stilde\Gamma,
\stilde\xi\}$ do not enter because they affect signal and
noise equally.

Even though $q_{\text{n}}(t)$ and $f_{\text{n}}(t)$ are
statistically independent, the above result is fully consonant
with predictions from field theory \cite{Buonanno:02b,
Buonanno:01a, Buonanno:01b, Buonanno:02a} of correlations
between ``shot noise'' and ``radiation-pressure noise.''  We
identify shot noise with fluctuations in $y_{\text{n}}(t)$ and
radiation noise with the sum
$f_{\text{n}}(t)+f_{\text{c}}(t)$.  Then from
Fig.~\ref{fig:block diagram} we see that the combined effect
of the measurement kernel ${\xi}$ and the backaction kernel
$\sprime{\Gamma}$ is to apply a fluctuating radiation force
$f_{\text{c}}(t)$ that is deterministically related to the
shot noise $y_{\text{n}}(t)$.  This is the path integral
mechanism that correlates shot noise with radiation-pressure
noise.  Note that $f_{\text{c}}(t)$ does not enter into the
total strain noise~(\ref{eq:total strain noise}) because it is
already accounted by the spectral density $S_{q}(\omega)$ via
the feedback kernels $\xi$ and $\sprime\Gamma$ in
Fig.~\ref{fig:block diagram}.

Figure~\ref{fig:noise performance} shows the resulting
performance of the reference design.  As predicted by the Bode
diagram of Fig.~\ref{fig:PIMA_bode}, the sensitivity of the system
is greatly amplified in the frequency band where mechanical
and optical resonances coincide, \emph{viz.}, the
band~\mbox{10--30~Hz}.  From Table~\ref{tab:connections} we
find that the process noise $S_{f}(\omega)$ and measurement
noise $S_{q}(\omega)$ satisfy a device-independent equality
\begin{equation}
	\label{eq:reciprocity}
	S_{f}(\omega) S_{q}(\omega) = \hbar^{2}/4.
\end{equation}
Minimizing $S_{h}^{\text{tot}}(\omega)$ subject to this
equality defines the sprung mass quantum
limit $S_{h}^{\text{\,sprung}}(\omega)$:
\begin{align}
	\label{eq:SQL}
	S_{h}^{\text{\,sprung}}(\omega)& \myedef
	\underset{\scriptstyle \{S_{f},S_{q}\}}{\text{min}}
	S_{h}^{\text{tot}}(\omega)  
	\notag \\
	& = \frac{\hbar}{m^{2}\omega^{4} L^{2} |\stilde{G}(\omega)|} 
\end{align}
where the dynamical kernel $\stilde{G}(\omega)$ is held fixed
during minimization.  It follows that
$S_{h}^{\text{\,sprung}}(\omega)$ sets a rigorous lower bound
to the interferometer noise $S_{h}^{\text{tot}}(\omega)$:
\begin{equation}
	\label{eq:sprung limit}
   S_{h}^{\text{tot}}(\omega) \ge  S_{h}^{\text{\,sprung}}(\omega).
\end{equation}
This lower bound is included in Fig.~\ref{fig:noise
performance} and is seen to be saturated at two discrete
frequencies: \mbox{$\sim 15\ \text{Hz}$} and \mbox{$\sim 25\
\text{Hz}$}.  Physically speaking, at these frequencies the
process noise $S_{f}$ and the measurement noise $S_{q}$ are
optimally balanced for strain detection.

\input{PIMA_figure_5.txt}

As a final check, the assumption of free test mass dynamics yields
the standard quantum limit (SQL)
\begin{align}
	\label{eq: free mass SQL}
	S_{h}^{\text{SQL}}(\omega) & \myedef \lim_{1/|\stilde{G}|\to
m\omega^{2}} S_{h}^{\text{tot}}(\omega)  \notag \\
& \myecheck \frac{\hbar}{m \omega^{2} L^{2}}
\end{align}
in accord with the literature \footnote{%
The literature value is $S_{h}^{\text{SQL}}(\omega) = 8
\hbar/(m_{\text{mir}} \omega^{2} L^{2})$.  This agrees with
(\ref{eq: free mass SQL}) upon taking $m_{\text{mir}}\to 4
m_{\text{mode}}$, with $m_{\text{mode}}$ the motional mass of
a four-mirror interferometer, then $m_{\text{mode}}\to m$, with
$m$ our single-mirror mass, and finally inserting a factor of
1/2 to convert to a two-sided spectral density.}.

As shown in Fig.~\ref{fig:noise performance}, the reference
design beats the SQL in the 10--30~Hz band.  But this does not
signify any evasion of the rigorous quantum limits
(\ref{eq:reciprocity}--\ref{eq:sprung limit}), because the SQL
assumption of free test mass dynamics is not justified,
\emph{viz.}, the dynamical kernel $\stilde{G}(\omega)$ differs
greatly from the free kernel $-1/(m\omega^{2})$ in consequence
of optical forces and springs.

These results illustrate a fundamental principle of field
theory: all physics can be derived from the scattering matrix. 
In our case the scattering matrix is the optical amplitude
(\ref{eq:FPa}--\ref{eq:FPb}), and from this sole input the
path integral/measurement amplitude formalism constructs both
the classical dynamics and the quantum noise.

\section{Accord with the Gravity Wave Detection Literature}
\label{sec: literature}
\noindent Now we will show that the path integral results of
the preceding section---both dynamical and
noise-related---accord with prior results from the
gravity-wave detection community \cite{Caves:80, Caves:81,
Braginsky:02, Braginsky:01c, Braginsky:98, Buonanno:02b,
Buonanno:01a, Buonanno:01b, Buonanno:02a, Braginsky:01,
Khalili:01, Braginsky:99}, which were obtained mainly by
operator and field-theoretic methods.

Showing accord is daunting because the gravity-wave community
has a tradition---extending back at least twenty years---of
``lively controversies'' \cite{Caves:80}.

Much recent discussion has been stimulated by the work of
Braginsky and colleagues \cite{Braginsky:01}, who have 
been cited \cite{Buonanno:01a} as showing that ``the test-mass
wave-function aspect of the uncertainty principle is
irrelevant to the operation of a [gravity-wave]
interferometer''.  

To put a sharp point on the issue, how can the path
integral/measurement amplitude formalism, in which the test
mass is explicitly quantized but light is not, be equivalent
to other---seemingly opposite---formalisms in which the test
mass is not explicitly quantized but light is?

In showing that there need be no contradiction, we will build
upon a seminal article by Caves \cite{Caves:80} and an
analysis of photodetection by Gardiner and Zoller
\cite{Gardiner:00}.  Caves' analysis revealed ``two different,
but equivalent points of view regarding the origin of \ldots\
radiation-pressure fluctuations.''  Seeking further equivalent
points of view, we can identify in Gardiner and Zoller's
analysis (and in much other quantum optics literature) at
least six variables in which quantum fluctuations occur
(Table~\ref{table:variables}).

\input{PIMA_table_4.txt}

Viewed as field operators, these variables are linked by
Maxwell's equations, such that fluctuations in any one
operator determine the fluctuations of all the others (up to
boundary conditions on the optical field).  Thus, any one of
these variables can reasonably serve as the focus of an
``equivalent point of view'' in the sense of Caves.

The path integral/measurement amplitude method amounts to a
point of view that is focussed upon the test mass
trajectory $q(t)$ and can be formalized as follows:
\begin{PIMAitemize}
	\item The quantum dynamics of the mirror are embodied (\ref{eq: path integral}--b) 
	in a path integral over $q(t)$.
	\item Maxwell's equations are enforced by the optical kernel
	(\ref{eq:classical scattering amplitude}), with $q(t)$
	as the source term.
	\item Optical boundary conditions are
	specified via the measurement amplitude 
	(\ref{eq:fundamental}).
\end{PIMAitemize}
Adherants of this point of view can nonetheless consistently
agree with the very different point of view of Caves' articles
\cite{Caves:80, Caves:81}, which focus upon vacuum
fluctuations at the input ports.  In a measurement amplitude
formalism the port fluctuations appear implicitly in the
photon counting statistics of the measurement
amplitude~(\ref{eq:fundamental}), in accord with Gardiner and
Zoller's dictum \cite{Gardiner:00}:
\begin{quote}
	Under the conditions that normally apply for a practical
	photodetector, the `out' electron field has the same
	statistics as the `in' photon field.
\end{quote}
Thus, vacuum fluctuations entering at the input port
necessarily appear in the photon statistics.  This
reconciles---at least in principle---the port-oriented point
of view with the path integral/measurement amplitude formalism.

This suggests the general principle that a path
integral/measurement amplitude analysis should agree with any
other analysis that treats at least one variable quantum
mechanically, enforces Maxwell's and Newton's equations, and
imposes compatible boundary conditions on the optical fields.

As a test of this ``many viewpoints'' principle, we
have carried through a path integral analysis of each of the
measurement schemes that were analyzed in \cite{Braginsky:01}
by operator methods; we find exact agreement between the two
formalisms in all cases.  This work will be reported in a
separate article.  However, a conceptual issue arose in which
the language of control theory proved more precise than the
language of physics; this precision played a key role in
reconciling the two viewpoints.

The issue is: what is a free mass?  To a control engineer the
question is ill-posed, because an appropriate control kernel
can create dynamics that are equivalent to a virtual
spring, even though no physical spring is present. 
Conversely, a physical spring attached to a test mass can be
veiled by a control kernel, such that the controlled dynamics
are equivalent to those of a free mass.

Veiled springs pose a conceptual obstacle in measurement
theory because they allow violation of the free-mass standard
quantum limit by test masses that are only seemingly free.  We
found in \cite{Braginsky:01} several examples of meters
that---upon computing an equivalent measurement amplitude,
path integral, and control diagram---proved
quantum-mechanically equivalent to a physical spring plus a
spring-veiling controller.  In every case the reciprocity
relation (\ref{eq:reciprocity}) was satisfied, such that
violations of the standard quantum limit were due to the
veiled spring.

\section{Conjectured Limits to Interferometric
Test Mass Observation}
\noindent We set forth in this section conjectured limits that, if
correct, constrain all designs for interferometric gravity
wave detectors, including past and future quantum
nondemolition designs.  To maintain an intellectual
equilibrium---and to help sustain the gravity wave community's
tradition of ``lively controversy''---we will suggest in
Section~\ref{sec:Discussion} several approaches by which these
conjectures might be proved wrong or evaded.

To start, we propose the following conventional definition of
a free mass:
\begin{definition}
	\label{th:definition 1}
A free mass
is defined to have a transfer function
$\stilde{\lcal{G}}(\omega) \myedef
\stilde{q}(\omega)/\stilde{f}(\omega) = -1/(m\omega^{2}) = 1/(ms^{2})$.
\end{definition}
Here $s$ is the Laplace variable traditionally preferred over
$\omega$ by control engineers, and $\stilde{\lcal{G}}(\omega)$
includes both mechanical and optical springs.  This definition
unveils hidden springs, and from a control engineering point
of view is the most natural definition.

We then propose the following conjecture, which formalizes the
path integral result (\ref{eq:reciprocity}):
\begin{conjecture}
	\label{th:conjecture 1}
	For any stationary lossless interferometric measurement
	processes, the measurement noise spectral density
	$S_{q}(\omega)$ and the process noise spectral density
	$S_{f}(\omega)$ satisfy an exact equality
	\begin{equation}
		\label{eq:conjecture}
	S_{q}(\omega)S_{f}(\omega) = \hbar^{2}/4 
	\end{equation}
	and these noise processes are
	uncorrelated.
\end{conjecture}
Here the ``stationary'' constraint excludes stroboscopic
\cite{Braginsky:99} and squeezed \cite{Rugar:91} measurements,
and ``lossless'' is understood to mean ``no unobserved
decoherence.''  We recall from the discussions following
(\ref{path:b}) and (\ref{eq:total strain noise}) that the
measurement noise $q_{\text{n}}(t)$ and process noise
$f_{\text{n}}(t)$ are not equivalent to shot noise and
radiation-pressure noise, but rather are mnemonic aids whose
sole role is to remind us to include $S_{f}$ and $S_{q}$
in~(\ref{path:b}).

The point of Conjecture~\ref{th:conjecture 1} is to suggest
that (\ref{eq:conjecture}) need not be regarded as a limit to
be approached, but instead provides us with an exact law of
nature that even the most clumsily designed experiments
cannot violate, provided only that all decoherence is
observed and no observations are discarded.  This viewpoint
facilitates the informatic investigations we propose in
Section~\ref{sec:Discussion}.

Definition~\ref{th:definition 1} and
Conjecture~\ref{th:conjecture 1} lead immediately to a lemma
(derived in (\ref{eq:reciprocity}--\ref{eq:SQL})):
\begin{lemma}
	\label{th:lemma 1}
	For any stationary interferometric test mass measurement, 
	the spectral density $S_{h}^{\text{tot}}(\omega)$ of 
	the equivalent gravitational strain noise satisfies
	an inequality
	\begin{equation}
		\label{eq:lemma}
	S_{h}^{\text{tot}}(\omega) \ge \frac{\hbar}{m^{2}\omega^{4} L^{2} |\stilde{G}(\omega)|}
	\end{equation}
	where $m$ is the reduced mirror mass, $L$ is the arm length,
	$\omega$ is the angular observation frequency, and
	$\stilde{G}(\omega)$ is the test mass transfer function, including optical forces.
\end{lemma}
Since this inequality---the sprung mass quantum limit---holds
even for squeezed photon detection statistics (as discussed
following (\ref{eq:fundamental})), the point of
Lemma~\ref{th:lemma 1} is to suggest the strong hypothesis
that \emph{all} stationary measurement schemes for exceeding
the standard quantum limit, if analyzed from the path
integral/measurement amplitude point of view, and with care
taken to unveil hidden springs, are equivalent to the design
strategy of the previous section, which can be formalized as
follows:
\begin{PIMAitemize}
\item Install an optical or mechanical spring that increases
$|\stilde{G}(\omega)|$ relative to the free mass value of
$1/(m\omega^{2})$, 
\item 
Simultaneously tune the sideband response of the cavity to
balance $S_{q}(\omega)$ and $S_{h}(\omega)$ such that the
sprung mass limit (\ref{eq:lemma}) is saturated over the
broadest feasible bandwidth, and
\item
Install a control kernel $\stilde\Gamma(\omega)$ to quench any
optomechanical instabilities and---if desired---alter or
veil the dynamical effects of the optical spring.
\end{PIMAitemize}
The nondemolition meters proposed in \cite{Braginsky:01} are
consistent with this strategy, but there are many other
proposed meters in the literature that remain to be considered
before it could be considered general.

If Lemma~\ref{th:lemma 1} is correct, then optimizing the
sensitivity of interferometric gravity wave detectors is a
problem that can be posed purely in terms of classical
optomechanical design.  Because the theoretical and practical
limits to maximizing $\stilde{G}(\omega)$ are not known,
Lemma~\ref{th:lemma 1} imposes no fundamental limit---quantum
or classical---on the sensitivity of interferometric gravity
wave detection.

\section{Discussion}
\label{sec:Discussion}
\noindent Conjecture~\ref{th:conjecture 1} and
Lemma~\ref{th:lemma 1} are suggested by the path
integral/measurement amplitude formalism, but they are far
from proved.  We will now outline a program by which they
might be proved wrong or evaded.  Beyond its intrinsic
scientific interest, this program would advance at least three
practical goals: quantum cryptography, single-spin imaging,
and interferometric gravity-wave detection.

Quantum cryptography and quantum entanglement considerations
arise naturally when we consider how to generalize the
measurement amplitude (\ref{eq:fundamental}) to the case of
multiple output ports.  A natural $n$-port \emph{ansatz} is
\begin{align}
	\label{eq:generalization}
\exp\left[
\frac{\lcal{M}(\{r_{i}(t)\},q(t))}{i\hbar}\right]
= & \mbox{\raisebox{-0.5ex}[0pt][0pt]{$\underset{i\in 1,n}{\mbox{\huge$\Pi$}}$}}
\Bigg(\bigg[\frac{a_i(t|q(t))}{|a_i(t|q(t))|}\bigg]^{r_{i}(t)}\notag \\ 
& \hspace*{-5em} \times \exp\bigg[
\frac{-(r_{i}(t)-|a_i(t|q(t))|^2)^2}
{4 \alpha_{i}|a_i(t|q(t))|^2}\bigg] \Bigg)
\end{align}
where $\{a_{i}(t|q(t)),r_{i}(t),\alpha_{i}\}$ specify the
amplitude functional, detection rate, and photon count
squeezing at the $i$'th output port.  This measurement
amplitude---or a similar expression---would have to be
rigorously grounded in field theory before
Conjecture~\ref{th:conjecture 1} could be regarded as a
theorem.  Furthermore, the design analysis of real-world
gravity-wave detectors, whether analytically or by quantum
numerical simulation, also requires an explicit $n$-port
measurement amplitude.

The following thought experiment suggests how challenging such
a field-theoretic grounding might be.  Consider a four-port
interferometer observing a single test mass, in which Alice
monitors Ports~1 and~2 while Bob monitors Ports~3 and~4.

Alice and Bob decide---independently and secretly---how to
process their ports.  For example, Alice can decide to count
photon rates $|a_{1}|^{2}$ and $|a_{2}|^{2}$, or alternatively
she can count $|a_{1}- i a_{2}|^{2}/2$ and $|a_{1}+ i
a_{2}|^{2}/2$; Alice's data records will in general be very
different depending on her choice, as will her inferred values
of $S_{q}(\omega)$ and $S_{f}(\omega)$.

Alice's choices must be invisible to Bob, and Bob's choices
must be invisible to Alice; otherwise causality is violated. 
But depending on the quantum dynamics of the test mass, there
is at least the possibility of quantum entanglement of Alice
and Bob's port amplitudes.  Furthermore, Alice and Bob have
the option---at least in principle---of storing their light
away, for analysis at some future time by a method to be
decided later; such delayed choices must also be mutually
transparent.  

Such thought experiments suggest that rigorously justifying or
refuting (\ref{eq:conjecture}--\ref{eq:generalization}) will
encompass nontrivial issues of quantum entanglement,
consistent with a recent proposal by Marshall \emph{et al.}
\cite{Penrose:02}.

Quantum entanglement issues appear with redoubled subtlety when we
consider magnetic resonance force microscopy (MRFM).  As with
interferometric gravity wave detection, MRFM experiments
monitor test masses by optical interferometry
\cite{Sidles:92,Rugar:92,Sidles:92b}.  Their differing
physical scale---nanograms, nanowatts, and nanometers for
MRFM interferometers versus kilograms, kilowatts, and
kilometers for gravity wave interferometers---is not
particularly relevant to the physics.  More fundamentally
different is the MRFM community's goal of observing the
non-classical force signal from an individual spin.

Early work in MRFM included the test mass quantum dynamics,
but did not include a quantum analysis of the measurement
process \cite{Sidles:92b}.  Conversely, direct interferometric
observation of a single spin has been analyzed
\cite{Sidles:96xxx}, but this ``toy'' analysis did not include
any test mass dynamics.  Thus, no integral quantum analysis of
a combined interferometer/spin/test-mass system is available
at present.  As MRFM technology approaches attonewton force
sensitivity \cite{Stowe:97}, such that long-envisioned
single-spin detection and bioimaging applications
\cite{Sidles:92b} approach feasibility, this fundamental quantum
measurement challenge is gaining in urgency.

The statistical nature of the transition between spin-up and
spin-down signals has crucial practical significance for the
MRFM community; it strongly conditions the design of optimal
signal processing algorithms.  This is is a practical
embodiment of a decades-old question: when and how does a
quantum wave function collapse?

Similarly gaining in urgency is the practical challenge of how
best to tune and operate gravity wave interferometers.  For
want of theoretical guidance, the interferometer of
Section~\ref{sec:Example} was tuned empirically.  Had we sent
the output photons to Alice---per the discussion above---along
with a homodyne reference, Alice might have achieved much
better sensitivity, \emph{viz.}, a substantially more optimal
balance between $S_{q}(\omega)$ and $S_{f}(\omega)$.

Alice's secretly improved sensitivity has to be transparent to
Bob's simultaneous observation.  Thus, part of Alice and Bob's
communication challenge is to agree on how best to establish a
consensus test mass trajectory, and how best to distinguish
shot noise from radiation-pressure noise in their combined
data records.  The resulting Alice-Bob dialog would cast new
light on these contentious issues---doubly so if they were
sharing nonclassical spin signals in an MRFM context.

In summary, the quantum signal processing and control
challenges in both gravity wave interferometry and magnetic
resonance force microscopy are mathematically well-posed,
reasonably accessible via the path integral/measurement
amplitude formalism, rich in fundamental physics and
unexplored information-theoretic issues, and rich in quantum
system engineering challenges.  A new generation of
instruments based on these technologies---instruments of
unprecedented sensitivity, if they can be made to
work---promises to open new worlds for scientific observation
and exploration.
\begin{acknowledgments}
This work was supported by the National Institutes of Health,
the National Science Foundation, and the Defense Advanced
Research Projects Agency's MOSAIC Program.  The author thanks
Dan Rugar of IBM for asking ``How does the Stern-Gerlach
effect really work?''  Doug Cochran, Alfred Hero, and Karoly
Holzer of the MOSAIC program pointed out the practical
importance of single-spin signal transitions in MRFM. Kip
Thorne extended the hospitality of the LIGO group, and
Alessandra Buonanno illuminated the gravity-wave detection
literature in many helpful conversations.
\end{acknowledgments}
\bibliography{PIMA_prd,masterMRFMbiblio}
\end{document}

%% file: PIMA_figure_1.txt
\begin{figure}[b]
\vspace*{-2ex}%
\hspace*{-2ex}%
\includegraphics[width = 3.5in]{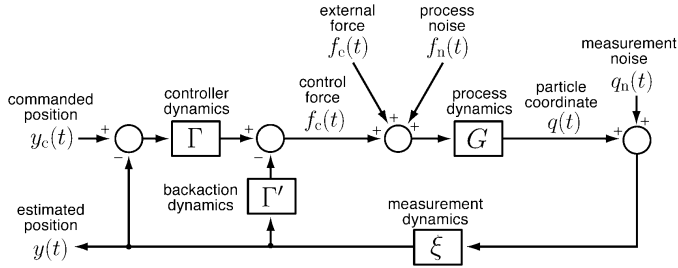}%
\hspace*{-2ex}%
\vspace*{-2ex}%
\caption{\label{fig:block diagram}
	The control theory block diagram associated with equations
	\mbox{(\ref{eq:control a}--c)}, or equivalently the path
	integral (\ref{eq: path integral}--b).
}
\end{figure}

%% file: PIMA_table_1.txt
\begin{table*}[t]
\caption{\label{tab:functionals}
Table of functional densities that appear in (\ref{eq: path
integral}).  See Table~\ref{tab:connections} for the
connection between the kernels $\{\Gamma,\sprime\Gamma,\theta,\psi,\xi\}$ of these
densities and the system dynamics specified by
Eqs.~(\ref{path:a}--b).} 
\begin{tabular}{r@{$\,=\,$}l@{$\ \Leftrightarrow\ $}l}
	\hline\hline
\multicolumn{3}{l}{\myStrut\textbf{Lagrangian test mass action}}\\
\multicolumn{3}{c}{}\\[-1ex]
	 $\lcal{L}(q(t))$ 
	 	& $\displaystyle \frac{\sdot{q}^{2}(t)}{2 m} - \frac{1}{2} m \omega_{0}^{2} q^{2}(t)$ 
		& $\left\{\text{\parbox{\myboxwidth}{\raggedright The Lagrangian of a test particle
		of mass $m$ with (optional) spring constant $m\omega_{0}^{2}$.}}\right.$\\
\multicolumn{3}{l}{\myStrut\textbf{External force and control functionals}}\\
\multicolumn{3}{c}{}\\[-1ex]
		$\lcal{H}_{\text{f}}(q(t),f_{\text{e}}(t))$ 
	 	& $\displaystyle -q(t) f_{\text{e}}(t)$ 
		& $\left\{\text{\parbox{\myboxwidth}{\raggedright Generates the dynamical effects
		of the external \\ force $f_{\text{e}}(t)$.}}\right.$\\
\multicolumn{3}{l}{}\\[-1ex]
	 $\lcal{H}_{\Gamma}(q(t),y(t),y_{\text{c}}(t))$ 
	 	& $\displaystyle -q(t)\!\!\int_{-\infty}^{\infty}\!\!\!d\sprime{t}\, \Gamma(t-\sprime{t}) 
		(y_{\text{c}}(\sprime{t})-y(\sprime{t}))$ 
		& $\left\{\text{\parbox{\myboxwidth}{\raggedright Generates the control force appropriate to
    the commanded position $y_{\text{c}}(t)$.}}\right.$\\
\multicolumn{3}{l}{\myStrut\textbf{Measurement functionals}}\\
	 $\lcal{M}_{\sprime\Gamma}(q(t),y(t))$ 
	 	& $\displaystyle +q(t)\!\!\int_{-\infty}^{\infty}\!\!\!d\sprime{t}\, \sprime\Gamma(t-\sprime{t}) y(\sprime{t})$ 
		& $\left\{\text{\parbox{\myboxwidth}{\raggedright Generates a
		fluctuating backaction force that
		is deterministically correlated with fluctuations in $y(t)$.}}\right.$\\
\multicolumn{3}{c}{}\\[-1ex]
		$\lcal{M}_{\theta}(q(t),b(t))$ 
	 	& $\displaystyle  -b(t)\!\!\int_{-\infty}^{\infty}\!\!\!d\sprime{t}\, \theta(t-\sprime{t}) q(\sprime{t})$  
		& $\left\{\text{\parbox{\myboxwidth}{\raggedright Generates a backaction force
		parameterized by the deterministic function $b(t)$.}}\right.$\\
\multicolumn{3}{c}{}\\[-1ex]
	 $\lcal{M}_{\psi}(q(t),r(t))$ 
	 	& $\displaystyle + \frac{1}{2} \int_{-\infty}^{\infty}\!\!\!d\sprime{t}\,dt''\, \psi(t-t',t-t'') q(\sprime{t})q(t'')$ 
		& $\left\{\text{\parbox{\myboxwidth}{\raggedright Generates a frequency-dependent 
		backaction \\ spring constant.}}\right.$\\
		\multicolumn{3}{c}{}\\[-1ex]
		$\hspace*{1em}\lcal{M}_{\xi}(q(t),y(t),r(t))$ & $\displaystyle -\frac{\hbar}{4\gamma^{2}} \Big(y(t)-
		\!\!\int_{-\infty}^{\infty}\!\!\!d\sprime{t}\,\xi(t-\sprime{t})
		q(\sprime{t})\Big)^{2}$ &
		$\left\{\text{\parbox{\myboxwidth}{\raggedright Correlates
		the measured value $y(t)$ with the test \\ mass position
		$q(t)$.}}\right.$\\
\multicolumn{3}{c}{}\\[-1ex]
\hline
\hline
\end{tabular}
\end{table*}

%% file: PIMA_table_2.txt
\begin{table}[t]
\setlength{\myboxwidth}{1.1in}
\caption{\label{tab:connections} The
connection between the measurement kernels
$\{\psi,\theta,\xi\}$ of Table~\ref{tab:functionals} and the
system dynamics and noise specified by Eqs.~(\ref{path:a}--b) and
Fig.~\ref{fig:block diagram}.}
\begin{tabular}{r@{$\,=\,$}l@{$\ \Leftrightarrow\ $}l}
	\hline\hline
\multicolumn{3}{l}{\myStrut\textbf{Backaction dynamical effects}}\\
\multicolumn{3}{c}{\hspace*{-1ex}\hspace*{\columnwidth}}\\[-1ex]
	 $\stilde{G}^{-1}(\omega) $ & $m (\omega_{0}^{2}-\omega^{2}) + 
	 \stilde{\psi}(\omega,-\omega)$ &
	 $\left\{\text{\parbox{\myboxwidth}{\raggedright dynamical
	 kernel.}}\right.$\\
\multicolumn{3}{c}{}\\[-1ex]
	 $\stilde{f}_{\text{e}}(\omega) $ 
	 & $\stilde{b}(\omega) \stilde{\theta}(-\omega)$ 
	 & $\left\{\text{\parbox{\myboxwidth}{\raggedright 
	 backaction force.}}\right.$\\
\multicolumn{3}{l}{\myStrut\textbf{Measurement noise PSDs}}\\
\multicolumn{3}{c}{}\\[-1ex]
	 $S_{q}(\omega) $ 
	 & $\displaystyle \frac{\gamma^{2}}{|\xi(\omega)|^{2}}$ 
	 & $\left\{\text{\parbox{\myboxwidth}{\raggedright measurement noise.}}\right.$\\
\multicolumn{3}{c}{}\\[-1ex]
	 $S_{f}(\omega) $ 
	 & $\displaystyle \frac{\hbar^{2}}{4 S_{q}(\omega)}$
	 & $\left\{\text{\parbox{\myboxwidth}{\raggedright force noise.}}\right.$\\
\multicolumn{3}{c}{}\\[-1ex]
\hline
\hline
\end{tabular}
\end{table}

%% file: PIMA_table_3.txt
\begin{table}[t]
\setlength{\myboxwidth}{1.20in}
\caption{\label{tab:sidebandrules} 
Rules connecting the photon sideband amplitudes $\{\alpha,\beta\}$ 
to the measurement amplitudes 
$\{\psi,\theta,\xi,\Gamma\}$ of 
Tables~\ref{tab:functionals} and~\ref{tab:connections} for the
interferometer configuration of Fig.~\ref{fig:PIMA interferometer}}
\begin{tabular}{r@{$\,=\,$}l@{$\ \Leftrightarrow\ $}l}
\hline\hline
\multicolumn{3}{l}{\myStrut\textbf{Fields and variables}}\\
\multicolumn{3}{c}{\hspace*{-1ex}\hspace*{\columnwidth}}\\[-1ex]
	 $\stilde{y}(\omega)$ & $\displaystyle \stilde{r}(\omega)-2 \pi r_{\text{in}} \delta(\omega)$
	 &$\left\{\text{\parbox{\myboxwidth}{\raggedright photon flux\\ fluctuations.}}\right.$\\
\multicolumn{3}{c}{}\\[-1ex]
	 $\stilde{b}(\omega) $ 
	 & $2 \pi r_{\text{in}} \delta(\omega)$
	 &$\left\{\text{\parbox{\myboxwidth}{\raggedright 
	 light pressure.}}\right.$\\
\multicolumn{3}{c}{}\\[-1ex]
	 $\gamma^{2}$ & $\displaystyle \alpha_{\text{s}}r_{\text{in}}$
	 &$\left\{\text{\parbox{\myboxwidth}{\raggedright Sets the 
	 flux spectral density $S_r = \alpha_{\text{s}}r_{\text{in}}$.}}\right.$\\
\multicolumn{3}{l}{\myStrut\textbf{Measurement kernels}}\\
\multicolumn{3}{c}{}\\[-1ex]
	 $\sprime{\stilde{\Gamma}}(\omega) $ & $\displaystyle
	 -\frac{\hbar}{2 i}
	 \big[\stilde{\alpha}(-\omega)-\stilde{\alpha}^{\star}(\omega)\big]$
	 &$\left\{\text{\parbox{\myboxwidth}{\raggedright backaction control kernel.}}\right.$\\
\multicolumn{3}{c}{}\\[-1ex]
	 $\stilde{\theta}(\omega) $ 
	 & $-\sprime{\stilde{\Gamma}}(-\omega)$
	 &$\left\{\text{\parbox{\myboxwidth}{\raggedright 
	 couples light pressure to the test mass.}}\right.$\\
\multicolumn{3}{c}{}\\[-1ex]
	 $\stilde{\psi}(\omega) $ 
	 & $ -\hbar r_{\text{in}}\,$\parbox[t]{1in}{\raggedright $\text{Im}\big[\,2\stilde{\beta}(\omega,-\omega)$\\
	 \quad $-{\stilde\alpha}(\omega){\stilde\alpha}(-\omega)\big]$}
	 &$\left\{\text{\parbox{\myboxwidth}{\raggedright 
	frequency-dependent spring constant.}}\right.$\\
\multicolumn{3}{c}{}\\[-1ex]
	 $\stilde{\xi}(\omega) $ 
	 & $\displaystyle r_{\text{in}}\big[{\stilde\alpha}(\omega)+{\stilde\alpha}^{\star}(-\omega)\big]$
	 &$\left\{\text{\parbox{\myboxwidth}{\raggedright 
	 measurement kernel.}}\right.$\\
\multicolumn{3}{c}{}\\[-1ex]
\hline
\hline
\end{tabular}
\end{table}

%% file: PIMA_figure_2.txt
\begin{figure}[t]
\includegraphics[width = 2.9in]{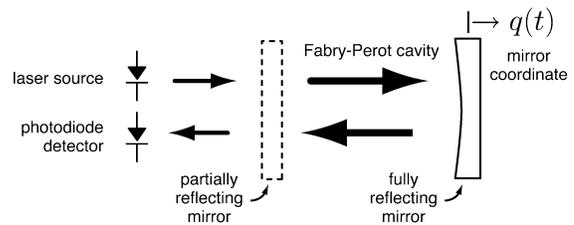}
\caption{\label{fig:PIMA interferometer}
	A Fabry-Perot cavity, with the end mirror serving as
	a test mass with coordinate $q(t)$.}
\end{figure}

%% file: PIMA_figure_3.txt
\begin{figure}[t]
\includegraphics[width = \columnwidth]{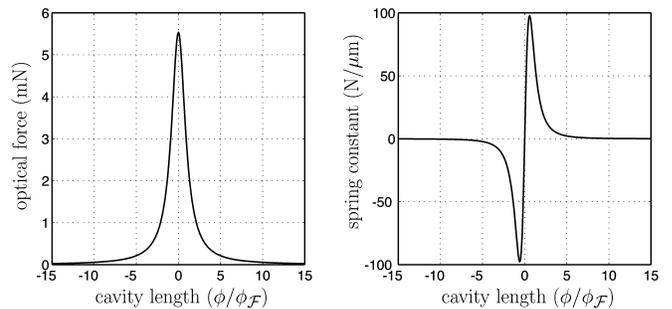}
\caption{\label{fig:statics}
	Static force and spring constant generated by the
	intra-cavity light.  Here $\phi \myedef k L + \pi$, with $L$
	the cavity length and $k$ the optical wavenumber, and
	$\phi_{\lcal{F}}$ is defined in (\ref{eq:finesse phase}).  }
\end{figure}

%% file: PIMA_figure_4.txt
\begin{figure}[t]
\includegraphics[width = \columnwidth]{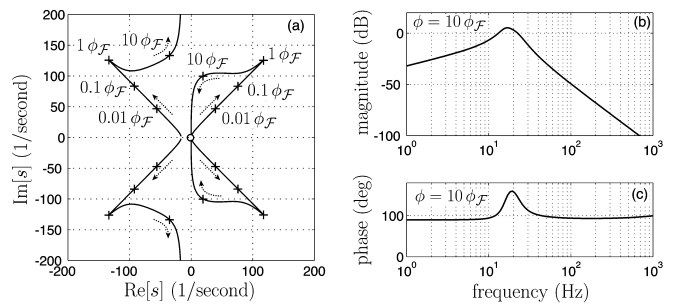}
\caption{\label{fig:PIMA_bode}
	Dynamical behavior of the test mass.  (a) Poles of the
	transfer function $\stilde{T}(\omega)$ as the cavity tuning
	is varied over $0<\phi<\infty$. The specific tunings
	$\phi/\phi_{\lcal{F}} \in \{0.01,0.1,1,10\}$ are marked
	with a `\mbox{%
	\protect\rule[0.5ex]{1ex}{0.15ex}%
	\protect\hspace{-1ex}%
	\protect\hspace{0.425ex}%
	\protect\rule[0.075ex]{0.15ex}{1ex}%
	\protect\hspace{0.4ex}%
	}'.
	A fixed zero at $s=0$ is marked with a
	`\protect\raisebox{-0.1ex}{\mbox{$\gsb\circ$}}'. 
	(b-c)~A~Bode plot of $\stilde{T}(\omega)$ evaluated for the
	particular cavity tuning $\phi = 10\,\phi_{\lcal{F}}$.  The
	magnitude axis is $20\,\log_{10}|\stilde{T}(2\pi f)\
	\text{N}/\mu\text{W}|$.}
\end{figure}

%% file: PIMA_figure_5.txt
\begin{figure}[t]
\includegraphics[width = \columnwidth]{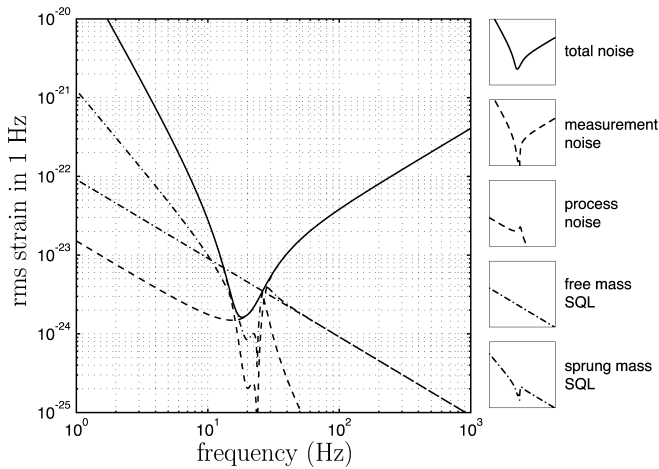}
\caption{\label{fig:noise performance}
	Noise performance of the Fabry-Perot cavity.  The rms total
	strain noise is defined to be $[2S_{h}^{\text{tot}}/(1\
	\text{second})]^{1/2}$, and all the other noise densities are
	normalized similarly.  The equivalent one-sided bandwidth is
	one hertz.  }
\end{figure}

%% file: PIMA_table_4.txt
\begin{table}[t]
\begin{ruledtabular}
\caption{
\label{table:variables}
Variables in which quantum fluctuations
are commonly identified.
}
\begin{tabular}{c}
\parbox{3in}{
\vspace*{-2ex}
\begin{align}%
	\lb{j}_{\text{in}}(t)&:\ \text{the laser source current,}\notag\\
	\lb{A}_{\text{in}}(t)&:\ \text{the gauge field at the input port(s),}\notag\\
	\lb{A}_{\text{cav}}(t)&:\ \text{the gauge field within the cavity,}\notag\\
	\lb{j}_{\text{mir}}(t)&:\ \text{the mirror current,}\notag\\
	\lb{A}_{\text{out}}(t)&:\ \text{the gauge field at the output port(s),}\notag\\
	\lb{j}_{\text{out}}(t)&:\ \text{the photodiode sink.}\notag
	\end{align}%
}
\vspace*{-2ex}
\end{tabular}
\end{ruledtabular}
\end{table}